\documentclass{biostatistics}

\usepackage{amsmath}
\usepackage{amssymb}
\usepackage{verbatim}
\usepackage{graphicx,psfrag,epsf}
\usepackage{multirow}
\usepackage{natbib}
\usepackage{booktabs}
\usepackage{multirow}
\usepackage{verbatim}
\usepackage{url} 
\usepackage{array}

\DeclareMathOperator*{\argmax}{arg\,max}

\title[Delayed Treatment Effects and Single Crossings]{Nonparametric Analysis of 
Delayed Treatment Effects using Single-Crossing Constraints}

\begin{document}

\author{Nicholas C. Henderson$^{1\ast}$, Kijoeng Nam$^2$ and Dai Feng$^3$\\[4pt]
$^{1}$Department of Biostatistics, University of Michigan, Ann Arbor, MI, USA
\\[2pt]
$^{2}$BARDS, Merck \& Co., Inc., North Wales, PA, USA \\ [2pt]
$^{3}$Data and Statistical Sciences, AbbVie Inc., North Chicago, IL, USA \\ [2pt]
{*email: nchender@umich.edu}}

\markboth%
{N.C. Henderson \& K. Nam \& D. Feng}
{Delayed Treatment Effects with Single-Crossing Constraints}

\maketitle

\begin{abstract}
{Clinical trials involving novel immuno-oncology (IO) therapies frequently exhibit survival profiles which violate the proportional hazards assumption due to a delay in treatment effect, and in such settings, the survival curves in the two treatment arms may have a crossing before the two curves eventually separate. To flexibly model such scenarios, we describe a nonparametric approach for estimating the treatment arm-specific survival functions which constrains these two survival functions to cross at most once without making any additional assumptions about how the survival curves are related. A main advantage of our approach is that it provides an estimate of a crossing time if such a crossing exists, and moreover, our method generates interpretable measures of treatment benefit including crossing-conditional survival probabilities and crossing-conditional estimates of restricted residual mean life. We demonstrate the use and effectiveness of our approach with a large simulation study and an analysis of reconstructed outcomes from a recent combination-therapy trial.}{censored data; clinical trial; constrained estimation; immuno-oncology; non-proportional hazards}
\end{abstract}

\section{Introduction}
\label{sec:intro}

Recent advances in immuno-oncology (IO) therapies for the treatment of cancer have led 
to the development of a variety of treatments which show great potential 
for improving long-term patient outcomes. While very promising, patient response to such immunotherapies is often quite different when compared to more traditional cytotoxic agents such as chemotherapy.  
Indeed, it is well-recognized that IO drugs frequently exhibit a clear delay in treatment effect when they are compared with 
standard chemotherapies, and moreover, the nature of the delay in this treatment effect is often such that
the estimated survival curves in the two treatment arms have a crossing at some time point after randomization. 
Due to this feature of immunotherapies, traditional comparisons between IO drugs and chemotherapies can have a number of limitations.

For time-to-event endpoints in randomized clinical trials, the log-rank test is the conventional choice for testing the superiority or non-inferiority of an active treatment over a control, and whenever the log-rank test passes a threshold for statistical significance, a hazard ratio is typically reported as one of the chief measures of treatment efficacy. 
However, in settings with survival curve crossings where the proportional hazards assumption is plainly violated, the interpretation of both the log-rank test and an estimated hazard ratio from a Cox proportional hazards model could be unclear. A variety of alternatives to the hazard ratio have been suggested in the context of evaluating the efficacy of immunotherapies or other context where the proportional hazards assumption regularly fails.
These include, for example, difference in restricted mean survival time (RMST) (\cite{zhao2016}, \cite{pak2017}),
difference in milestone survival (\cite{chen2015}), average hazard ratios (\cite{schemper2009}), and the proportion
of patients that are functionally ``cured" (\cite{chen2013}). In the context of testing
the equality of the arm-specific survival curves under possible delayed treatment effects, a number of weighted log-rank tests have been suggested including, for example, 
the piecewise weighted log-rank test (APPLE) proposed in \cite{xu2017} and weighted log-rank tests from the 
the Fleming-Harrington family which can place more weight on later time points (\cite{harrington1982}, \cite{rahman2019}). 
Combination tests which combine multiple weighted log-rank test statistics have also been proposed
for analyzing IO trials (\cite{lin2020}).

While the measures of treatment effect listed above have a clear interpretation in the absence of proportional hazards and reporting such measures of treatment efficacy can certainly be valuable, 
imposing estimation constraints on how the survival curves may cross can provide additional information by which to evaluate a treatment that exhibits delayed benefit. 
To this end, we propose nonparametric estimates of the treatment-specific survival 
functions that allow for at most one crossing of the two survival functions without requiring that the crossing time be pre-specified. 
The approach we describe for estimating the survival curves under a single-crossing constraint can be thought of as a two-stage procedure. In the first stage, one finds conditional estimates of the survival function by maximizing a nonparametric log-likelihood that is conditional 
on the value of two crossing parameters. Then, in the second stage, one estimates these crossing parameters by maximizing a profile log-likelihood function. This produces estimates of the two survival curves which satisfy the single-crossing constraint, and it generates an estimate of the crossing time and an estimate of which survival curve is initially dominant.
While flexible semiparametric approaches allowing for the crossing of treatment-specific 
survival functions have been proposed (see, e.g., \cite{yang2005} and \cite{demarqui2019}), our approach makes no assumptions about how the treatment-specific survival curves are related other than that they cross at most once, and moreover,
our approach provides a likelihood-based estimate of the crossing time if a positive crossing time is determined to be present.

One of the advantages of imposing a single-crossing constraint is that
situations where there is a delay in treatment effect and improved long-term survival in the active treatment arm are often better modeled as having a single crossing after which point the survival curve in the active treatment arm consistently dominates the survival curve of the control arm. Such patterns of delayed treatment effect have been observed in many recent immuno-oncology trials, and hence, constrained modeling in such trials has the potential to improve estimation performance and provide greater interpretability when comparing the two arm-specific estimates of the survival functions. 
An important advantage of using single-crossing constraints is that it yields estimates and uncertainty intervals for the time that the crossing occurs. 
The estimate of the crossing-time parameter can be useful in 
both assessing when the active treatment begins to show superiority
and as an interpretable measure that can be used as a component of additional measures of treatment efficacy that are designed for scenarios with delayed treatment effect which we outline in detail in Section \ref{sec:inference}. These efficacy measures include the proportion of patients surviving up to crossing, crossing-time conditional restricted residual mean life, crossing-time conditional survival probabilities, and pre/post-crossing average hazard ratios.

This paper is organized as follows.  
In Section \ref{sec:nonpara_model}, we describe a two-stage nonparametric estimation procedure which assumes that the treatment arm-specific survival curves cross at most once but otherwise makes no additional assumptions about the forms of the survival functions, and we briefly outline several potential extensions of this approach. Section \ref{sec:inference} discusses a number of estimands of interest that can capture relevant concerns about the treatment impact in cases where survival curves may cross, and we describe how our method can be used to estimates these terms. Section \ref{sec:inference} also describes hypothesis tests of interest under delayed treatment effect. Section \ref{sec:simulations} shows the results from a simulation study which evaluates the estimation performance of our method across six piecewise-exponential simulation settings.
Section \ref{sec:dataexample} shows an application of our method to a recent trial involving a novel combination immunotherapy in the treatment of non-small-cell lung cancer, and Section \ref{sec:discussion} concludes with a brief discussion.

\vspace{-.1in}

\section{Nonparametric Estimation for Survival Curves with Single Crossings} 
\label{sec:nonpara_model} 

\subsection{Survival Curve Profiles and Notation}
We assume that $n$ patients have been enrolled in a randomized clinical trial consisting of two treatment arms. The primary outcome is the time to some event of interest, and for the $i^{th}$ individual in the study, we let $T_{i}$ denote the time-to-failure for this event of interest. Instead of directly observing $T_{i}$, we observe the follow-up time $Y_{i} = \min\{ T_{i}, C_{i} \}$ and the event indicator $\delta_{i} = I( T_{i} \leq C_{i})$ where $C_{i}$ denotes the censoring time, and $I(\cdot)$ denotes the indicator function. 
We let $A_{i} = 1$ denote that patient $i$ was assigned to the active treatment arm, and we let $A_{i} = 0$ denote that patient $i$ was assigned to the control arm. We also assume that censoring is noninformative in the sense that the terms from the censoring distribution can be factored out of the likelihood function (\cite{lawless2011}), and we assume that $C_{i}$ and the treatment arm assignment $A_{i}$ are independent.

\begin{figure}
\centering
     \includegraphics[width=5.5in,height=4.75in]{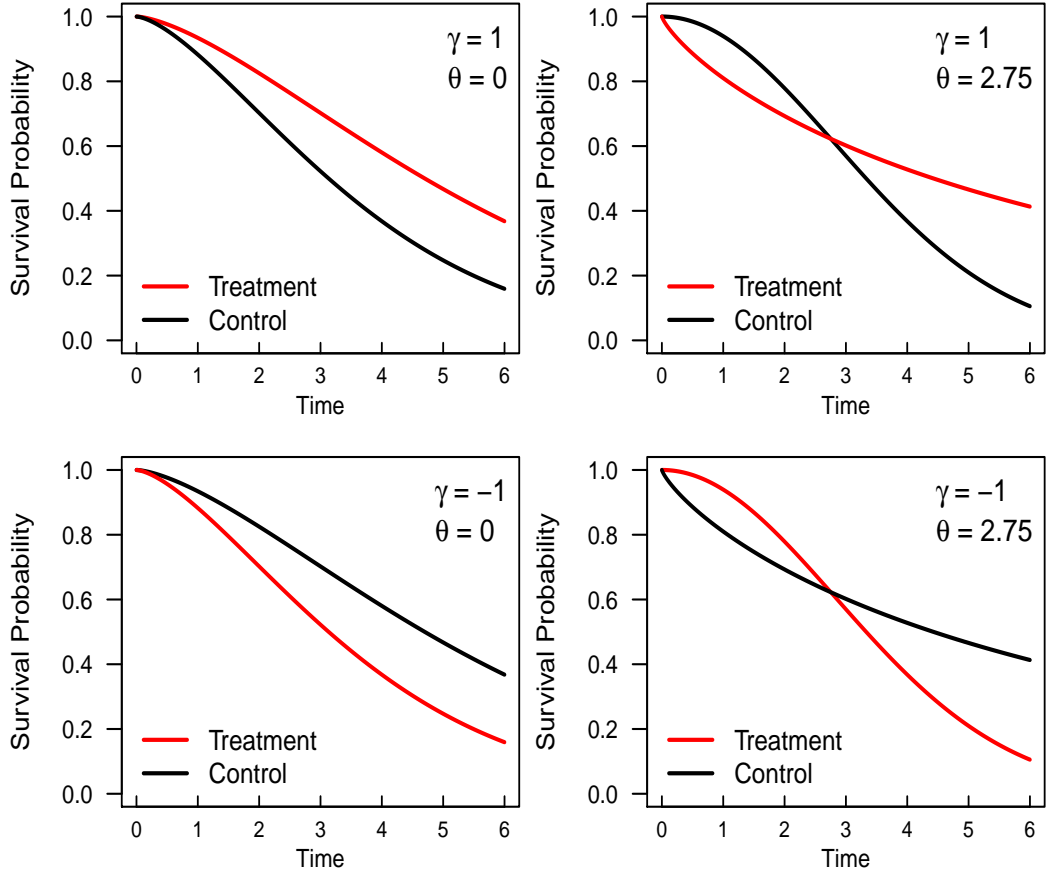}
\caption{Four possible survival profiles. The crossing parameters $\theta$ and $\gamma$ refer 
to the crossing-time parameter and initial dominance parameter respectively. We set $\theta = 0$ if
no crossing occurs. A value of $\gamma = 1$ indicates that the control treatment arm is dominant
before the crossing while a value of $\gamma = -1$ indicates that the active treatment arm is
dominant before the crossing.}
\label{fig:four_profiles}
\end{figure}

We let $S_{a}(t) = P(T_{i} > t|A_{i} = a)$ denote the survival function for those patients assigned to treatment group $a \in \{0, 1\}$. We consider an analysis where these survival functions are allowed to exhibit four distinct ``survival profiles'' according to whether or not a survival curve crossing occurs. Specifically, we allow for the possibility that the two survival functions $S_{0}(t)$ and $S_{1}(t)$ ``cross", but we limit the number of crossings so that they can occur at most once.
The crossing restriction implies that we will have one of four survival profiles, where each profile refers to a distinct crossing pattern of the survival functions. These four possible survival profiles are as follows: (1) a survival profile where treatment arm $a=1$ completely dominates $a=0$, namely $S_{1}(t) \geq S_{0}(t)$ for all $t \geq 0$; (2) a survival profile where $a=0$ dominates $a=1$ before some crossing time $\theta$ but $a=1$ dominates $a=0$ afterwards, i.e., $S_{0}(t) \geq S_{1}(t)$ for $0 \leq t \leq \theta$ and $S_{1}(t) \geq S_{0}(t)$ for $t > \theta$; (3) a survival profile where $a=1$ dominates $a=0$ before some crossing time $\theta$ but $a=0$ dominates $a = 1$ afterwards; and (4) a survival profile where $a = 0$ completely dominates $a = 1$. Figure \ref{fig:four_profiles} illustrates each of these four possible survival profiles. 

To enforce one of the four possible profiles implied by Figure \ref{fig:four_profiles}, we introduce the two crossing parameters $\theta \geq 0$ and $\gamma \in \{-1, 1\}$.
The parameter $\theta$ represents the crossing time of the survival functions, and the discrete parameter $\gamma$ determines which treatment arm has the initially dominant survival function. Specifically, when $\gamma=1$ the survival functions are assumed to exhibit the following behavior 
\begin{equation}
    S_{0}(t) \geq S_{1}(t) \quad \textrm{ for } 0 \leq t \leq \theta \qquad \textrm{ and } \qquad S_{0}(t) \leq S_{1}(t) \quad \textrm{ for } t > \theta, \label{eq:gamma1_condition}
\end{equation}
and when $\gamma = -1$, the survival functions are assumed to obey the following inequalities
\begin{equation}
    S_{0}(t) \leq S_{1}(t) \quad \textrm{ for } 0 \leq t \leq \theta \qquad \textrm{ and } \qquad S_{0}(t) \geq S_{1}(t) \quad \textrm{ for } t > \theta. \label{eq:gammaneg1_condition}
\end{equation}
If there is a $\theta > 0$ such that either constraints (\ref{eq:gamma1_condition}) or constraints (\ref{eq:gammaneg1_condition}) 
hold, we will say that $\theta$
is a non-trivial \textit{crossing time} of the survival functions $S_{0}$ and $S_{1}$. If there is no such $\theta > 0$,
we must either have $S_{0}(t) \geq S_{1}(t)$ or $S_{1}(t) \geq S_{0}(t)$ for all $t \geq 0$ in which case we set $\theta$ to the trivial crossing time $\theta = 0$.

When constructing estimates of the survival functions, we can only enforce the infinite-dimensional constraints
implied by (\ref{eq:gamma1_condition}) and (\ref{eq:gammaneg1_condition}) at a finite number of time points. 
In our implementation, we enforce the survival function constraints (\ref{eq:gamma1_condition}) and (\ref{eq:gammaneg1_condition})
at each of the observed event times. To this end, we let $0 < t_{1} < t_{2} < ... < t_{m}$ denote the unique, ordered event times from both treatment arms, and hence, $m$ represents the total number of events occurring in either of the treatment arms. 
For $\gamma = 1$ and a fixed value of $\theta$, we enforce the following constraints
\begin{eqnarray}
S_{0}( t_{j} ) &\geq& S_{1}(t_{j}) \qquad \textrm{ for all $j$ such that } t_{j} \leq \theta \nonumber \\
S_{0}( t_{j} ) &\leq& S_{1}(t_{j}) \qquad \textrm{ for all $j$ such that } t_{j} > \theta. 
\label{eq:gamma1_Hcondition}
\end{eqnarray}
Similarly, for $\gamma = -1$ and a fixed value of $\theta$, we enforce the following constraints 
\begin{eqnarray}
S_{0}( t_{j} ) &\leq& S_{1}(t_{j}) \qquad \textrm{ for all $j$ such that } t_{j} \leq \theta \nonumber \\
S_{0}( t_{j} ) &\geq& S_{1}(t_{j}) \qquad \textrm{ for all $j$ such that } t_{j} > \theta. 
\label{eq:gammaneg1_Hcondition}
\end{eqnarray}
Note that (\ref{eq:gamma1_Hcondition}) and (\ref{eq:gammaneg1_Hcondition}) together imply that we will have a different set of constraints for each choice of $(\theta, \gamma)$. In the next subsection, we describe 
our procedure for estimating the arm-specific survival functions when both $\theta$ and $\gamma$ are assumed to be fixed.

\subsection{Estimating the Survival Functions with Fixed Crossing Parameters} \label{ss:fixed_cross}

We first describe nonparametric estimation of the survival functions $S_{a}(t)$ assuming both $\theta$ and $\gamma$ are known.
As detailed in \cite{park2012} in the context of finding the constrained nonparametric maximum likelihood estimate of survival functions under a stochastic ordering constraint, the constrained maximum likelihood estimates of the survival functions will be discrete with potential jumps only at the event times $t_{1}, \ldots, t_{m}$. Because of this, we define $u_{ja}(\theta, \gamma)$ as the jump of $\log S_{a}(t)$ at the point $t_{j}$ when it is assumed that the true crossing-time parameter is $\theta$ and the true initial dominance parameter is $\gamma$. More specifically, 
$u_{ja}(\theta,\gamma) = \log\{ S_{a}(t_{j}) \} - \log\{ S_{a}(t_{j}-) \}$,
where $S_{a}( t- )$ denotes the left limit of $S_{a}( t )$ at time $t$. The term $h_{ja}(\theta, \gamma) = 1 - \exp\{ u_{ja}(\theta, \gamma) \} = 1 - S_{a}(t_{j})/S_{a}(t_{j}-)$ can be interpreted as the discrete hazard at time point $t_{j}$. Following \cite{park2012} and \cite{johansen1978}, the log-likelihood function to be maximized in this context is
\begin{equation}
\log L\{ \mathbf{u}_{0}(\theta,\gamma), \mathbf{u}_{1}(\theta, \gamma) \} = \sum_{a=0}^{1}\sum_{j=1}^{m} \Big[ d_{ja}\log\{ 1 - \exp(u_{ja}(\theta,\gamma))\} + (R_{ja} - d_{ja})u_{ja}(\theta,\gamma) \Big], 
\label{eq:np_log_lik}
\end{equation}
where $\mathbf{u}_{0}(\theta,\gamma)$ and $\mathbf{u}_{1}(\theta,\gamma)$ are the $m \times 1$ vectors having components $u_{j0}(\theta, \gamma)$ and $u_{j1}(\theta, \gamma)$ respectively. In (\ref{eq:np_log_lik}), $R_{j0} = \sum_{i=1}^{n} (1 - A_{i})I(Y_{i} \geq t_{j})$ denotes the number of individuals at risk in the control arm at time $t_{j}$, and $R_{j1} = \sum_{i=1}^{n} A_{i} I(Y_{i} \geq t_{j})$ denotes the number of individuals at risk in the active treatment arm at time $t_{j}$.
In (\ref{eq:np_log_lik}), $d_{j0} = \sum_{i=1}^{n} (1 - A_{i})\delta_{i}I(Y_{i} = t_{j})$ denotes the number of events in the control arm at time $t_{j}$ while $d_{j1} = \sum_{i=1}^{n} A_{i}\delta_{i}I(Y_{i} = t_{j})$ denotes the number of events in the active treatment arm at time $t_{j}$.

In the absence of any crossing constraints on the survival functions, the only constraints on the vector $\mathbf{u}_{a}(\theta,\gamma)$ would be $u_{ja}(\theta, \gamma) \leq 0$, for $j = 1,\ldots, m$. 
Because the survival functions $S_{0}(t)$ and $S_{1}(t)$ associated with 
the vectors $\mathbf{u}_{0}(\theta, \gamma)$ and $\mathbf{u}_{1}(\theta, \gamma)$ have the form $S_{a}( t ) = \exp\big\{ \sum_{j=1}^{m} u_{ja}(\theta,\gamma) I(t_{j} \leq t) \big\}$, the crossing 
constraints (\ref{eq:gamma1_Hcondition}) and (\ref{eq:gammaneg1_Hcondition}) can be expressed as a collection of linear inequality constraints. Specifically, we can represent both the crossing constraints and the inequality constraints $u_{ja}(\theta, \gamma) \leq 0$ as
$[\mathbf{a}_{k}(\theta, \gamma)]^{T}\mathbf{u}(\theta, \gamma) \geq 0$ for $k = 1, \ldots, 3m$, where $\mathbf{a}_{k}(\theta, \gamma) \in \mathbb{R}^{2m}$ and $\mathbf{u}(\theta, \gamma) = \big( \mathbf{u}_{0}(\theta,\gamma)^{T}, \mathbf{u}_{1}(\theta,\gamma)^{T}\big)^{T}$. For $k \leq m$, the vectors $\mathbf{a}_{k}(\theta, \gamma)$ are constructed to enforce the crossing constraints (\ref{eq:gamma1_Hcondition}) or (\ref{eq:gammaneg1_Hcondition}). If we define $v(\theta) = \max\{ j: t_{j} \leq \theta \}$ if $\theta \geq t_{1}$ and $v(\theta) = 0$ if $\theta < t_{1}$ , then $\mathbf{a}_{k}(\theta, 1)$ for $k \leq m$ is given by 
\begin{equation}
\mathbf{a}_{k}(\theta, 1) 
= \begin{cases}
\big(\mathbf{1}_{k}, \mathbf{0}_{m - k}, -\mathbf{1}_{k}, \mathbf{0}_{m - k} \big), & \textrm{ if } k \leq v(\theta) \nonumber \\
\big(-\mathbf{1}_{k}, \mathbf{0}_{m - k}, \mathbf{1}_{k}, \mathbf{0}_{m - k} \big), & \textrm{ if } v(\theta) < k \leq m
\end{cases}
\end{equation}
where $\mathbf{1}_{l}$ denotes a vector of length $l$ containing all ones, $\mathbf{0}_{l}$ denotes a vector of length $l$ containing all zeros, and $\mathbf{0}_{0}$ denotes a vector of ``length 0" that should be ignored.
Similarly, $\mathbf{a}_{k}(\theta, -1)$ is given by
\begin{equation}
\mathbf{a}_{k}(\theta, -1) 
= \begin{cases}
\big(-\mathbf{1}_{k}, \mathbf{0}_{m - k}, \mathbf{1}_{k}, \mathbf{0}_{m - k} \big), & \textrm{ if } k \leq v(\theta) \nonumber \\
\big(\mathbf{1}_{k}, \mathbf{0}_{m - k}, -\mathbf{1}_{k}, \mathbf{0}_{m - k} \big), & \textrm{ if } v(\theta) < k \leq m
\end{cases}
\end{equation}
In order to enforce the constraints $u_{ja}(\theta, \gamma) \leq 0$,  $\mathbf{a}_{k}(\theta, \gamma)$ is, for any value of $(\theta, \gamma)$, defined as 
$\mathbf{a}_{k}(\theta, \gamma) = (\mathbf{0}_{k-m-1}, -1, \mathbf{0}_{3m - k})$ for $k = m+1, \ldots, 3m$.

The maximum likelihood estimates $\hat{\mathbf{u}}_{a}(\theta, \gamma)$ of $\mathbf{u}_{a}(\theta,\gamma)$ can be expressed as the solution to the following optimization problem that has linear inequality constraints
\begin{equation}
    \textrm{maximize } \log L\big\{ \mathbf{u}_{0}(\theta,\gamma), \mathbf{u}_{1}(\theta, \gamma) \big\} 
    \qquad \textrm{ subject to } \quad \mathbf{A}_{\theta, \gamma}\begin{bmatrix} \mathbf{u}_{0}(\theta, \gamma)^{T} \\ \mathbf{u}_{1}(\theta, \gamma)^{T} \end{bmatrix} \geq \mathbf{0}_{3m}, 
    \label{eq:optim_problem}
\end{equation}
where $\mathbf{A}_{\theta, \gamma}$ is the $3m \times 2m$ matrix whose $k^{th}$ row is $\mathbf{a}_{k}(\theta, \gamma)^{T}$.
The above optimization problem involves maximizing a concave function subject to linear inequality constraints, and hence, any local maximum is also guaranteed to be a global maximum (see e.g., \cite{boyd2004convex}). 
In our implementation, we use sequential quadratic programming (\cite{nocedal2006})
to compute the solution $\big( \hat{\mathbf{u}}_{0}(\theta, \gamma), \hat{\mathbf{u}}_{1}(\theta, \gamma) \big)$ of (\ref{eq:optim_problem}). Initialization of $\mathbf{u}_{0}(\theta, \gamma)$, $\mathbf{u}_{1}(\theta, \gamma)$ is done by minimizing, subject to the single-crossing constraint, the squared discrepancy $\sum_{a=0}^{1}\sum_{j=1}^{m} [u_{ja}(\theta, \gamma) - \log\{\hat{S}_{a}^{KM}(t_{j})/\hat{S}_{a}^{KM}(t_{j}-) \}]^{2}$,
where $\hat{S}_{a}^{KM}(t)$ is the Kaplan-Meier estimates of the survival function in treatment arm $a$. One possible limitation of this computational strategy is that sequential quadratic programming can become very computationally demanding for large values of $m$. 
One remedy for this is to group the follow-up times $Y_{1}, \ldots, Y_{m}$ into a collection of small ``bins" and set $Y_{i}$ to the midpoint of the bin to which it is assigned.

\subsection{Estimates of Crossing-Time Parameters $\theta$ and $\gamma$.}

The estimated vectors $\hat{\mathbf{u}}_{0}(\theta, \gamma)$ and $\hat{\mathbf{u}}_{1}(\theta, \gamma)$ will generate estimates of the two survival curves for fixed values of $(\theta, \gamma)$. To find the best values of the crossing parameters $(\theta, \gamma)$, we maximize the profile log likelihood function $\ell^{P}(\theta, \gamma)$ associated with  $\hat{\mathbf{u}}_{0}(\theta, \gamma)$ and $\hat{\mathbf{u}}_{1}(\theta, \gamma)$
\begin{equation}
\ell^{P}(\theta, \gamma) = \sum_{a=0}^{1}\sum_{j=1}^{m} \Big[ d_{ja}\log\{ 1 - \exp(\hat{u}_{ja}(\theta,\gamma))\} + (R_{ja} - d_{ja})\hat{u}_{ja}(\theta,\gamma) \Big].
\label{eq:profile_log_lik}
\end{equation}
We refer to the values $\hat{\theta}_{sc}$, $\hat{\gamma}_{sc}$ which maximize $\ell^{P}(\theta, \gamma)$ as the \textit{single-crossing constrained} estimates of the crossing parameters $\theta$ and $\gamma$.
Because the conditional estimates $\hat{\mathbf{u}}_{a}(\theta, \gamma)$ do not change as $\theta$ varies over each of the intervals $(t_{j-1}, t_{j})$ and can only change at each $t_{j}$, the single-crossing constrained estimates $\hat{\theta}_{sc}$ and $\hat{\gamma}_{sc}$ can be found by solving the following discrete optimization problem
\begin{equation}
(\hat{\theta}_{sc}, \hat{\gamma}_{sc})
= \argmax_{\theta \in \{0,t_{1}, \ldots, t_{m-1} \}, \gamma \in \{-1, 1\} }   \ell^{P}(\theta, \gamma). 
\label{eq:discrete_optim}
\end{equation}
The reason for only considering values of $\theta$ up to $t_{m-1}$ is because both $\theta = 0$ and $\theta = t_{m}$ refer to situations where one survival function dominates the other survival function at every time point $t_{1}, \ldots, t_{m}$. Thus, including $\theta = t_{m}$ as a possible crossing time is superfluous as $(\theta, \gamma) = (0, 1)$ and $(\theta, \gamma) = (0, -1)$ cover both scenarios where one survival curve dominates the other at each of the event times $t_{j}$.

The vectors $\hat{\mathbf{u}}_{a}(\hat{\theta}_{sc}, \hat{\gamma}_{sc})$ generate the following estimates of the survival functions
\begin{equation}
\hat{S}_{a}^{sc}(t) = \exp\Bigg\{ \sum_{j=1}^{m} \hat{u}_{ja}( \hat{\theta}_{sc}, \hat{\gamma}_{sc}) I(t_{j} \leq t)  \Bigg\}. \nonumber 
\end{equation}
We refer to $\hat{S}_{0}^{sc}(t)$ and $\hat{S}_{1}^{sc}(t)$ as the \textit{single-crossing constrained} estimates of the survival functions. Note that both $\hat{S}_{0}^{sc}(t)$ and $\hat{S}_{1}^{sc}(t)$ are flat for $t \geq t_{m}$, and hence if $\hat{S}_{0}^{sc}(t)$, $\hat{S}_{1}^{sc}(t)$ satisfy the single-crossing constraint over $[0, t_{m}]$, they will satisfy it for all time points.

\subsection{Alternative Single-Crossing Constraints}
The estimation strategy outlined in Sections 2.1-2.3 focuses on single-crossing constraints for the survival functions, but other related single-crossing constraints could potentially be incorporated using a similar approach.
We briefly mention a few interesting possible extensions below. While we explore non-smooth estimation of hazard functions with single-crossing constraints in our application in Section \ref{sec:dataexample}, we do not explore the other mentioned extensions further as they lie beyond the scope of this paper.

\noindent
\subsubsection{Non-smooth Estimation of Hazard Functions under Single-crossing Constraints} \label{ss:discrete_hazards}
In many cases, it is more sensible to place single-crossing constraints on the hazards rather than on the survival curves. In this context, without imposing any smoothness conditions we would want the discrete hazards $h_{ja}(\theta, \gamma) = 1 - S_{a}(t_{j})/S_{a}(t_{j}-)$ in one treatment arm to be larger (smaller) before some crossing time $\theta$ and remain smaller (larger) for $j$ such that $t_{j} > \theta$.
Because of the connection $h_{ja}(\theta,\gamma) = 1 - \exp\{ u_{ja}(\theta, \gamma) \}$ between the $u_{ja}(\theta, \gamma)$ and the discrete hazards $h_{ja}(\theta, \gamma)$, we can express the single-crossing constraints on the discrete hazards as
\begin{eqnarray}
u_{j0}(\theta, \gamma) &\geq& u_{j1}(\theta, \gamma) \qquad  \text{ for all $j$ such that } t_{j} \leq \theta \nonumber \\
u_{j0}(\theta, \gamma) &\leq& u_{j1}(\theta, \gamma) \qquad  \text{ for all $j$ such that } t_{j} > \theta, \label{eq:hazard_constraints}
\end{eqnarray}
if $\gamma = 1$ with both inequalities reversed whenever $\gamma = -1$. As in the case of estimating survival functions under single-crossing constraints, one would first, for fixed values of $(\theta, \gamma)$, find conditional maximum likelihood estimates of $u_{ja}(\theta, \gamma)$ by maximizing the log-likelihood (\ref{eq:np_log_lik}) subject to constraints (\ref{eq:hazard_constraints}). After this, one would find estimates of the crossing parameters $(\theta, \gamma)$ by maximizing the associated profile log-likelihood function (\ref{eq:profile_log_lik}).

\noindent
\subsubsection{Smooth hazard functions with single-crossing constraints} 
To find smoothly-estimated hazard functions, one can consider hazard functions $h_{a}(\cdot|\theta, \gamma)$ of the form
\begin{equation}
h_{a}(t|\theta,\gamma) = 1 - \exp\Big\{\sum_{j=1}^{m} s_{ja}(t) u_{ja}(\theta, \gamma) \Big\} = 1 - \exp\Big\{ \mathbf{s}_{a}(t)^{T}\mathbf{u}_{a}(\theta, \gamma) \Big\}, \label{eq:smooth_representation}
\end{equation}
for a choice of smoothing weights $\mathbf{s}_{a}(t) = \big(s_{1a}(t), \ldots, s_{m_{a}a}(t)\big)^{T}$.
A common choice of smoothing weights, for example, would be $s_{ja}(t) = \frac{1}{b}K\{ (t - t_{j})/b \}$ for some symmetric kernel function $K(\cdot)$ and bandwidth $b > 0$. Under formulation (\ref{eq:smooth_representation}), inequalities for the hazard functions at time points $t_{j}$ can be expressed as linear inequality constraints
of the form $\mathbf{s}_{1}(t_{j})^{T}\mathbf{u}_{1}(\theta, \gamma) \geq \mathbf{s}_{0}(t_{j})^{T}\mathbf{u}_{0}(\theta, \gamma)$, and hence, to compute smooth estimates of the hazard functions one could estimate the $\hat{u}_{ja}(\theta, \gamma)$ by maximizing the log-likelihood function (\ref{eq:np_log_lik}) subject to the linear inequality constraints implied by the form of the hazard functions in (\ref{eq:smooth_representation}). One would then find estimates of the crossing-time parameters $(\theta, \gamma)$ by maximizing the associated profile log-likelihood function (\ref{eq:profile_log_lik}).

An alternative to this approach would be to simply smooth estimated discrete hazards that have been found using the approach outlined in Section \ref{ss:discrete_hazards}. While this may work well in many situations, this approach would not guarantee that the smoothed hazard function estimates will satisfy the single-crossing constraint. 

\vspace{-8pt}
\noindent
\subsubsection{Covariate Adjustment}
Suppose each individual in the study has an additional covariate which we denote with $x_{i}$ for the $i^{th}$ individual. In this context, 
we would let $S_{a}(t|x_{i})$ denote the survival function conditional on being assigned to treatment arm $a$  and having covariate value $x_{i}$ and 
focus on crossing constraints for the ``baseline" survival functions $S_{0}(t)$ and $S_{1}(t)$ where $S_{a}(t) = S_{a}(t|0)$. If we define $u_{ja}(\theta, \gamma|x_{i}) = \log\{ S_{a}(t_{j}|x_{i})/S_{a}(t_{j}- |x_{i}) \}$ and $u_{ja}(\theta,\gamma) = u_{ja}(\theta,\gamma|0)$, a 
proportional hazards assumption with respect to the covariate $x_{i}$ would be
\begin{equation}
1 - \exp\{ u_{ja}(\theta,\gamma|x_{i}) \} = \big[ 1 - \exp\{ u_{ja}(\theta,\gamma) \} \big]\exp( \beta_{a}^{\theta,\gamma} x_{i} ), \nonumber  
\end{equation}
where $\beta_{a}^{\theta,\gamma}$ is an arm-specific regression coefficient that can depend on the values of the crossing parameters.
Note that this approach makes a proportional hazards assumption for the effect of the covariate $x_{i}$ within each treatment arm but does not make a proportional hazards assumption for the effect of the treatment arm assignment. Thus, this would still allow for the possibility of having a crossing between the two baseline survival functions $S_{0}(t)$ and $S_{1}(t)$. If we assumed that covariate-adjusted single-crossing constrained estimates of the baseline survival functions must have the same support as the overall single-crossing constrained estimates $\hat{S}_{a}^{sc}(t)$ - namely, support on the observed event times $t_{1}, \ldots, t_{m}$ - then to compute maximum likelihood estimates of $u_{ja}(\theta, \gamma)$ and $(\beta_{0}^{\theta,\gamma}, \beta_{1}^{\theta,\gamma})$ (for a fixed $(\theta, \gamma)$) in this context, we would maximize the following log-likelihood function
\begin{eqnarray}
&& \log L_{cov}\{ \mathbf{u}_{0}(\theta,\gamma),\mathbf{u}_{1}(\theta, \gamma), \beta_{0}^{\theta,\gamma}, \beta_{1}^{\theta,\gamma} \} = \sum_{a=0}^{1}\sum_{j=1}^{m} \sum_{i=1}^{n} \tilde{d}_{jai}\Big[\log\{ 1 - \exp(u_{ja}(\theta,\gamma))\} + \beta_{a}^{\theta,\gamma} x_{i} \Big] \nonumber \\
&+& \sum_{a=0}^{1}\sum_{j=1}^{m} \sum_{i=1}^{n} (\tilde{R}_{jai} - \tilde{d}_{jai})\log\Big\{ 1 - \big[1 - \exp\{ u_{ja}(\theta,\gamma) \} \big] \exp(\beta_{a}^{\theta,\gamma} x_{i}) \Big\}, \nonumber 
\end{eqnarray}
where $\tilde{d}_{jai} = \delta_{i}I(Y_{i} = t_{j})I(A_{i} = a)$ and $\tilde{R}_{jai} = I(Y_{i} \geq t_{j})I(A_{i} = a)$. The assumption that the covariate-adjusted single-crossing constrained estimates of $S_{a}(t)$ and the estimates $\hat{S}_{a}^{sc}(t)$ have the same support is similar to the assumptions made in, for example \cite{owen2001} and \cite{zhou2015}, in the empirical likelihood analysis of the Cox proportional hazards model. Because the relationship between $u_{ja}(\theta, \gamma)$ and the baseline survival functions in this context is the same as the relationship between $u_{ja}(\theta, \gamma)$ and the survival functions $S_{a}(t)$ in Sections 2.2-2.3, the constraints on the $u_{ja}(\theta, \gamma)$ required to impose a single-crossing constraint would be exactly the same as those in (\ref{eq:optim_problem}).

\vspace{-.1in}

\section{Model Inference} \label{sec:inference} 

\subsection{Estimands of Interest} \label{ss:estimands}

\textit{Milestone Survival Probabilities.} 
Comparing differences in estimated survival probabilities at one or several pre-specified time points $\hat{S}_{1}^{sc}(t_{1}^{*}) - \hat{S}_{0}^{sc}(t_{1}^{*}), \ldots, \hat{S}_{1}^{sc}(t_{q}^{*}) - \hat{S}_{0}^{sc}(t_{q}^{*})$
can be a useful way of characterizing the treatment effect over time without relying on any assumptions about proportional hazards. Because $\hat{\theta}_{sc}$
provides an estimate of precisely where the sign change in $S_{1}(t) - S_{0}(t)$ occurs, 
augmenting the survival probability differences at the milestones $t_{1}^{*}, \ldots, t_{q}^{*}$
with the estimated crossing time $\hat{\theta}_{sc}$ can be helpful when interpreting 
the estimated differences $\hat{S}_{1}^{sc}(t_{j}^{*}) - \hat{S}_{0}^{sc}( t_{j}^{*})$. 

\medskip
\noindent
\textit{The Proportion Surviving up to Crossing.}
The proportion surviving up to the crossing time in treatment arm $a$
is represented by the parameter $S_{a}( \theta )$. If both $S_{1}$ and $S_{0}$ are continuous, we will 
have $S_{1}(\theta) = S_{0}(\theta)$ and if either $S_{1}$ or $S_{0}$ are not continuous, these will be approximately equal as long as the true survival curves do not have large jumps. For this reason, we use $S_{a}(\theta)$ to denote the proportion surviving up to crossing, and in practice, we estimate this parameter with $\{\hat{S}_{1}^{sc}(\hat{\theta}_{sc}) + \hat{S}_{0}^{sc}(\hat{\theta}_{sc})\}/2$. The quantity $S_{a}( \theta)$ could be of particular interest if one is concerned about a substantial fraction of patients experiencing early events that occur before the crossing time. In these cases, reporting an estimate of $S_{a}( \theta )$ provides a measure of the fraction of patients who will survive long enough to reach the point at which the survival curve in the active treatment arm begins to dominate to the control-arm survival curve.

\medskip
\noindent
\textit{Restricted Mean Survival Time.}
The restricted mean survival time (RMST) (see, e.g., \cite{royston2013}) is defined as the expected time under follow-up for an individual assuming you only follow individuals up to some pre-specified time point $\tau$. Specifically, the RMST for individuals in treatment arm $a$ is defined as 
\begin{equation}
\textrm{RMST}_{a}(\tau) = E\{ \min( T_{i}, \tau ) | A_{i} = a \}. \nonumber
\end{equation}
$\textrm{RMST}_{a}(\tau)$ is equal to the area under the survival curve $S_{a}(t)$ between the time points $t = 0$ and $t = \tau$, and the difference $\textrm{RMST}_{1}( \tau ) - \textrm{RMST}_{0}( \tau )$ provides an interpretable measure of treatment effect regardless of whether or not the proportional hazards assumption holds. While providing an interpretable measure of treatment effect, the difference in RMST can mask important differences in survival that occur at earlier time points. One way of addressing this is to also examine differences in RMST for different choices of $\tau$ with differences in $\textrm{RMST}_{a}(\theta)$ perhaps being of key interest.

\medskip

\noindent
\textit{Restricted Residual Mean Life.} If one is interested in differences in survival for those that are longer survivors, the restricted residual mean life (RRML) function (see e.g., \cite{cortese2017}) is an appealing measure. The RRML function for treatment arm $a$ is defined at time $t$ as
\begin{equation}
\textrm{RRML}_{a}(t, \tau) = E\{ \min(T_{i}, \tau) - t| A_{i} = a, T_{i} \geq t\} 
= \int_{t}^{\tau} \frac{S_{a}(u)}{ S_{a}(t) } du. \nonumber
\end{equation}
The quantity $\textrm{RRML}_{a}(\theta, \tau)$ represents the expected on-study survival conditional on the fact that one has survived up to the crossing time $\theta$.

\medskip

\noindent
\textit{Crossing-Time Conditional Survival Curves.} In cases of delayed treatment where the two survival curves cross, it may be of interest to also plot survival probabilities conditional on surviving up to the point of crossing. Such conditional probabilities give the probability of surviving past a point of interest conditional on the fact that one has survived up to the crossing time. This conditional survival curve for patients in treatment arm $a$ is defined, for $t > \theta$, as
\begin{equation}
S_{a,cond}(t) = P\Big( T_{i} > t \Big| A_{i} = a, T_{i} > \theta \Big)
= S_{a}(t) \big/ S_{a}(\theta). \label{eq:conditional_survival_def}
\end{equation}
The conditional survival curves may be estimated directly using $\hat{S}_{a,cond}^{sc}( t ) = \hat{S}_{a}^{sc}(t)/\hat{S}_{a}^{sc}( \hat{\theta}_{sc})$. It is worth mentioning that $\textrm{RRML}_{a}(t, \tau) = \int_{\theta}^{\tau} S_{a,cond}( u ) du$.

\medskip

\noindent
\textit{Pre- and Post-crossing Average Hazard Ratios.} 
Comparing the average hazard ratios over the time periods before and after the 
crossing can provide an interpretable measure of treatment efficacy for 
longer survivors and can provide a good comparison for the relative improvement in treatment efficacy between earlier and later time points. Assuming arm-specific hazard functions $h_{a}(t)$ exist, we define, as in \cite{kalbfleisch1981}, the average hazard ratio using the
``active treatment-to-total" hazard ratio which measures the
average ratio between the active treatment-arm hazard $h_{1}(t)$
and the total hazard $h_{0}(t) + h_{1}(t)$ across time.
Specifically, for a truncation time $\tau$ and $\theta \in (0, \tau)$, we define the pre- and and post-crossing 
average hazard ratios as
\begin{equation}
\bar{\lambda}_{pre} = \frac{1}{\theta}\int_{0}^{\theta} \frac{h_{1}(t)}{h_{0}(t) + h_{1}(t) } dt \qquad \textrm{ and }\qquad \bar{\lambda}_{post} = \frac{1}{\tau - \theta} \int_{\theta}^{\tau} \frac{h_{1}(t)}{h_{0}(t) + h_{1}(t) } dt 
\label{eq:avghaz_parameters}
\end{equation}
respectively. One reason for using the ratio $h_{1}(t)/\{h_{0}(t) + h_{1}(t) \}$ rather than $h_{1}(t)/h_{0}(t)$ is to improve estimation stability as potentially very small estimated value of $h_{0}(t)$ could lead to highly variable estimates of $\bar{\lambda}_{pre}$ and $\bar{\lambda}_{post}$.
When assuming $\tau = t_{m}$, the parameters $\bar{\lambda}_{pre}$ and $\bar{\lambda}_{post}$ can be estimated by the following
quantities
\begin{eqnarray}
\hat{\bar{\lambda}}_{pre} &=& \frac{1}{\hat{\theta}_{sc}}\sum_{j=1}^{m} \frac{\hat{h}_{j1}(\hat{\theta}_{sc}, \hat{\gamma}_{sc})(t_{j} - t_{j-1})I(t_{j} \leq \hat{\theta}_{sc})  }{\hat{h}_{j0}(\hat{\theta}_{sc}, \hat{\gamma}_{sc}) + \hat{h}_{j1}(\hat{\theta}_{sc}, \hat{\gamma}_{sc}) } \nonumber \\
\hat{\bar{\lambda}}_{post} &=& \frac{1}{t_{m} - \hat{\theta}_{sc}}\sum_{j=1}^{m} \frac{\hat{h}_{j1}(\hat{\theta}_{sc}, \hat{\gamma}_{sc})(t_{j} - t_{j-1}) I(t_{j} > \hat{\theta}_{sc}) }{\hat{h}_{j0}(\hat{\theta}_{sc}, \hat{\gamma}_{sc}) + \hat{h}_{j1}(\hat{\theta}_{sc}, \hat{\gamma}_{sc}) }, \nonumber 
\end{eqnarray}
where $t_{0} = 0$ and $h_{ja}(\theta, \gamma) = 1 - \exp\{ u_{ja}(\theta, \gamma) \}$ is as defined in Section \ref{ss:fixed_cross}. Depending on the context, one could either use discrete hazard estimates $\hat{h}_{ja}(\hat{\theta}_{sc}, \hat{\gamma}_{sc})$ under the single-crossing constraints on the survival functions or the single-crossing constraints on the hazard function described in Section \ref{ss:discrete_hazards}.

\subsection{Hypothesis Testing}

While we can compute confidence intervals for certain parameters of interest that do not involve the crossing parameters $(\theta, \gamma)$, it can be useful to perform inference with respect to both $\theta$ and another parameter of interest $\phi$ (or collection of parameters) that represents a measure of treatment efficacy. For example, in traditional settings where it is assumed that proportional hazards hold, $\phi$ would frequently be a hazard ratio, but in settings with delayed treatment effect, choosing $\phi$ to be an alternative estimand such as difference in RMST may be more appealing. Combining $\theta$ and $\phi$ in a joint hypothesis test can address concerns about having a scenario where the estimated value of $\phi$ indicates overall treatment effectiveness, but substantial time elapses before the two survival curves clearly separate. Cases such as these may lead to concerns that most of the observed treatment benefit is mainly due to differences in long-term survivors.

In the aforementioned context, one possible hypothesis of interest is that both the efficacy parameter $\phi$ is sufficiently large
and the crossing time $\theta$ does not occur to late. This can be expressed more formally as
\begin{equation}
\mathcal{H}_{0}: \phi \leq \phi^{*} \textrm{ or } \theta \geq \theta^{*} \quad \textrm{ vs. } \quad \mathcal{H}_{A}: \phi > \phi^{*} \textrm{ and } \theta  < \theta^{*}, \label{eq:hyp_test1}
\end{equation}
where $\phi^{*}$ and $\theta^{*}$ are pre-specified values of $(\theta, \phi)$ which are determined to be clinically meaningful.  
Alternatively, if it is difficult to specify a time point before which the crossing should occur, one could instead require 
that survival in the active treatment arm should be sufficiently large whenever the crossing occurs. The hypothesis test of interest in this case would be
\begin{equation}
\mathcal{H}_{0}: \phi \leq \phi^{*} \textrm{ or } S_{1}(\theta) \leq p^{*} \quad \textrm{ vs. } \quad \mathcal{H}_{A}: \phi > \phi^{*} \textrm{ and } S_{1}( \theta ) > p^{*}. \label{eq:hyp_test2}
\end{equation}
One could test either (\ref{eq:hyp_test1}) or (\ref{eq:hyp_test2}) using a permutation test with the test statistics $\hat{\phi}$ and $\hat{\theta}_{sc}$ or $\hat{\phi}$ and $\hat{S}_{1}^{sc}( \hat{\theta}_{sc} )$ respectively. 
Another approach would be to test (\ref{eq:hyp_test1}) or (\ref{eq:hyp_test2}) by using a bootstrap procedure to construct one-sided confidence intervals for either the parameter $\eta_{1} = \min\{ \phi - \phi^{*}, \theta^{*} - \theta\}$ or the parameter $\eta_{2} = \min\{ \phi - \phi^{*}, S_{1}(\theta) - p^{*} \}$. If the lower bound of the confidence interval for $\eta_{1}$ is greater than zero, one would reject $\mathcal{H}_{0}$ in (\ref{eq:hyp_test1}).
Likewise, a lower bound for the confidence interval for $\eta_{2}$ greater than zero would imply that one should reject $\mathcal{H}_{0}$ in (\ref{eq:hyp_test2}).

\vspace{-.1in}

\section{Simulations} \label{sec:simulations}
\subsection{Estimation Performance with Piecewise Exponential Distributions}
\begin{figure}
\centering
     \includegraphics[width=5.5in,height=5in]{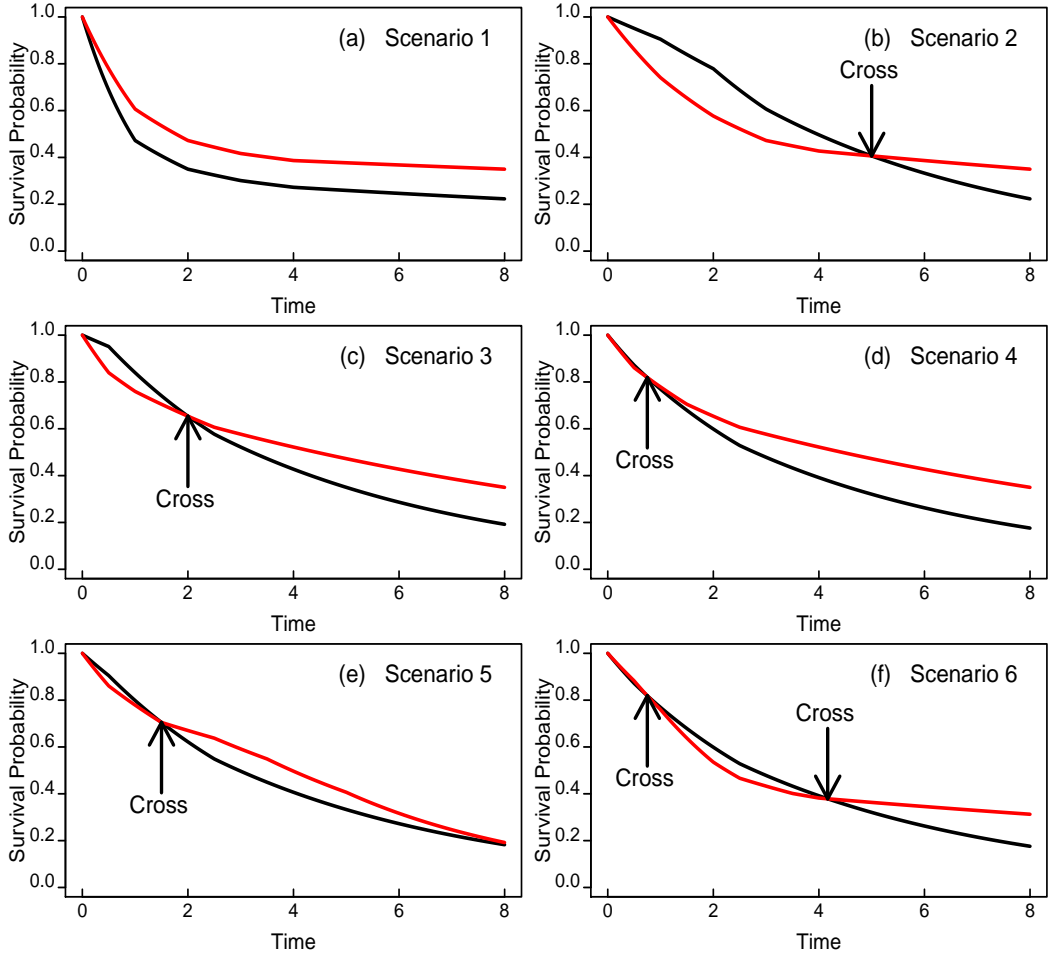}
\caption{Arm-specific survival curves from the six piecewise exponential simulation scenarios considered in Section \ref{sec:simulations}.}
\label{fig:pwexp_sim_settings}
\end{figure}

We considered six simulation scenarios where, in each scenario, it is assumed that survival follows a piecewise exponential distribution
in both treatment arms. The arm-specific survival curves for these six scenarios are depicted in Figure \ref{fig:pwexp_sim_settings}. The 
top-left graph in Figure \ref{fig:pwexp_sim_settings} (Figure 2(a)) depicts Scenario 1 where the survival curves never cross, and the survival curve for the active treatment arm always dominates the control-arm survival curve.
Figure \ref{fig:pwexp_sim_settings} (b) depicts Scenario 2 where there is a clear, unambiguous single crossing of the two survival curves at time point $5.0$. Figure \ref{fig:pwexp_sim_settings} (c) depicts Scenario 3 where the two survival curves have a single crossing near time point $2.0$.
While Scenario 3 has a single, distinct crossing, when compared with Scenario 2 the two survival curves in Scenario 3 do not have as much separation before the crossing occurs. 
Figure \ref{fig:pwexp_sim_settings} (d) shows the survival curves in Scenario 4
where there is a single crossing at time point $0.75$, but in this scenario, there is almost no separation between the curves before the crossing time. In Scenario $5$, there is a single crossing at time point $1.5$ with relatively little separation before the crossing and a diminishing treatment benefit that occurs towards the end of the time interval considered. In Scenario 6, there are two crossing times, but the later crossing is more ``distinct" than the first in the sense that the separation between the two curves is larger immediately before and after the crossing point.

For these simulations we set the total number of patients to $n = 200$, $n = 400$, and $n = 800$ with 
the number of patients split evenly between the two treatment arms for each choice of $n$.
For each of the six simulation scenarios and setting of $n$, we ran $200$ simulation replications. 
The censoring distribution used in each of the six scenarios was a uniform distribution from $4$ to $8$. While the percentage of survival outcomes which were observed event times varied across simulation scenarios, the percentage was between $55\%$ and $75\%$ for each of the six settings, and the percentage of observed events was typically $0 - 15\%$ larger in the control arm than in the active treatment arm. For each simulation setting, we evaluated the performance of the single-crossing constrained (SCC) procedure in estimating the following measures: difference in RMST at time point 7, the differences in the survival function at the time points 2 and 4, the crossing time $\theta$, the proportion surviving up to crossing $S_{a}(\theta)$, and the difference in RRML using the time points $\theta$ and $7$. 

Table \ref{tab:pwexp_sim_results} shows the mean-squared error (MSE) for the SCC estimates across the six piecewise-exponential simulation scenarios and the three choices of sample size. For parameters that do not involve the crossing time $\theta$, MSE was also computed for estimates based on the Kaplan-Meier estimates of the survival functions. As shown in this table, in scenarios with either no crossing or a distinct, single crossing (i.e., Scenarios 1-3), the SCC-based estimates had MSE performance which was consistently as good or better than the KM-based estimates for parameters for which such a comparison could be made. For example, in the $n=400$ settings, the reductions in MSE for the SCC-based estimator compared to the KM-based estimator of the RMST difference $\textrm{RMST}_{1}(7) - \textrm{RMST}_{0}(7)$ were $2.3\%$, $3.5\%$, and $2.2\%$ in Scenarios 1, 2, and 3 respectively, and
in the $n = 800$ settings the reductions in MSE for the RMST difference $\textrm{RMST}_{1}(7) - \textrm{RMST}_{0}(7)$ were $1.1\%$, $0.6\%$, and $0\%$ in Scenarios 1, 2, and 3 respectively.
MSE for the crossing-time estimator $\hat{\theta}_{sc}$ was lowest in both Scenarios 1 and 3 where there was either no crossing or an early, distinct crossing. The relatively poorer result for $\hat{\theta}_{sc}$ in Scenario 2 is likely due to the fact that, in this scenario, the true crossing time $\theta$ occurred much later in the study at a time where there would typically be much fewer individuals remaining in this study.
Despite this, the estimation performance for the estimator of $S_{a}(\theta)$ was quite good in Scenario 2 as both survival curves are much more flat towards the end of the study period. Scenario 5 was the one setting where estimation of the crossing-time parameter was notably poor. This was mainly due to the strong diminished treatment effect present in Scenario 5 which often resulted in 
a crossing-time estimate closer to the end of the considered time window rather than the much earlier true crossing time of 1.5. Estimation performance of the SCC-based estimators were overall quite poor in Scenario 6, but this was a scenario where the assumption of a single crossing was plainly violated.

\begin{table}[ht]
\centering
\begin{tabular}{cc rrrrrrrrr}
\toprule
&  & \multicolumn{2}{c}{{\small $\Delta \textrm{RMST}(7)$}} &
  \multicolumn{2}{c}{$S_{1}(2) - S_{0}(2)$} &
  \multicolumn{2}{c}{$S_{1}(4) - S_{0}(4)$} &
  \multicolumn{1}{c}{$\theta$} & \multicolumn{1}{c}{$S_{a}(\theta)$}
  & \multicolumn{1}{c}{\small $\Delta \textrm{RRML}(\theta, 7)$} \\
\cmidrule(r){3-4}\cmidrule(r){5-6}\cmidrule(r){7-8}\cmidrule(r){9-9}\cmidrule(r){10-10}\cmidrule(r){11-11}
n & Scenario & SCC & KM & SCC & KM & SCC & KM & SCC & SCC & SCC \\
\midrule
200 & 1 & 0.1348 & 0.1362 & 0.0043 & 0.0044 & 0.0037 & 0.0036 & 0.6502 & 0.0221 & 0.1548 \\ 
    & 2 & 0.1254 & 0.1289 & 0.0041 & 0.0042 & 0.0047 & 0.0046 & 3.0721 & 0.0326 & 0.2128 \\ 
    & 3 & 0.1369 & 0.1368 & 0.0046 & 0.0044 & 0.0050 & 0.0049 & 1.3836 & 0.0182 & 0.1639 \\ 
    & 4 & 0.1540 & 0.1381 & 0.0054 & 0.0048 & 0.0055 & 0.0050 & 2.2295 & 0.0342 & 0.1472 \\ 
    & 5 & 0.1890 & 0.1265 & 0.0065 & 0.0045 & 0.0071 & 0.0053 & 7.4537 & 0.0728 & 0.3152 \\ 
    & 6 & 0.2047 & 0.1323 & 0.0068 & 0.0045 & 0.0064 & 0.0048 & - & - & - \\ 
\midrule
400 & 1 & 0.0719 & 0.0736 & 0.0023 & 0.0024 & 0.0020 & 0.0020 & 0.3664  & 0.0092 & 0.0793 \\ 
    & 2 & 0.0669 & 0.0693 & 0.0019 & 0.0020 & 0.0025 & 0.0025 & 1.0848  & 0.0078 & 0.0728 \\ 
    & 3 & 0.0668 & 0.0683 & 0.0023 & 0.0023 & 0.0024 & 0.0024 & 0.6026  & 0.0088 & 0.0711 \\ 
    & 4 & 0.0730 & 0.0691 & 0.0024 & 0.0022 & 0.0025 & 0.0024 & 1.0826  & 0.0271 & 0.0595 \\ 
    & 5 & 0.1029 & 0.0636 & 0.0032 & 0.0022 & 0.0038 & 0.0025 & 7.4427  & 0.0699 & 0.1990 \\ 
    & 6 & 0.1184 & 0.0678 & 0.0035 & 0.0021 & 0.0036 & 0.0024 & -  & - & - \\
\midrule
800 & 1 & 0.0328 & 0.0332 & 0.0010 & 0.0010 & 0.0009 & 0.0009 & 0.0030 & 0.0008 & 0.0337 \\ 
    & 2 & 0.0317 & 0.0319 & 0.0009 & 0.0008 & 0.0011 & 0.0011 & 0.3767 & 0.0011 & 0.0249 \\ 
    & 3 & 0.0330 & 0.0330 & 0.0010 & 0.0010 & 0.0012 & 0.0012 & 0.2928 & 0.0043 & 0.0354 \\ 
    & 4 & 0.0360 & 0.0335 & 0.0012 & 0.0012 & 0.0012 & 0.0011 & 0.5016 & 0.0193 & 0.0278 \\ 
    & 5 & 0.0486 & 0.0299 & 0.0018 & 0.0012 & 0.0017 & 0.0011 & 5.3602 & 0.0459 & 0.1163 \\ 
    & 6 & 0.0610 & 0.0328 & 0.0022 & 0.0012 & 0.0017 & 0.0011 & - & - & - \\
\bottomrule
\end{tabular}
\caption{Mean-squared error (MSE) of single-crossing constrained (SCC) based estimators and Kaplan-Meier (KM) based estimators for several parameters from each of the six piecewise-exponential simulation scenarios. MSE for the Kaplan-Meier based estimators are
only shown when applicable. The parameters $\Delta RMST(7)$ and $\Delta RRML(\theta, 7)$ are defined as $\Delta RMST(7) = \textrm{RMST}_{1}(7) - \textrm{RMST}_{0}(7)$ and $\Delta RRML(\theta, 7) = \textrm{RRML}_{1}(\theta, 7) - \textrm{RRML}_{0}(\theta, 7)$ } 
\label{tab:pwexp_sim_results}
\end{table}

\vspace{-.2in}

\section{Data Example} \label{sec:dataexample}
In this section, we examine reconstructed survival outcomes from a recently completed 
phase 3 trial (\cite{hellmann2019}) examining the efficacy of a combination of immune checkpoint inhibitors, nivolumab plus ipilimumab, for the treatment of non-small-cell lung cancer. In this trial, patients were assigned to one of three treatments arms: a combination arm where nivolumab plus ipilimumab was administered, a monotherapy arm where nivolumab alone was administered, and a control arm where only chemotherapy was given. 
The primary endpoint in this study was overall survival (OS) in the combination therapy arm versus the chemotherapy arm
in the subpopulation of patients whose tumors had an expression level of the
programmed death ligand 1 (PD-L1) that was at least $1\%$. Among the group of patients
who had a PD-L1 expression of $1\%$ or more, $396$ patients were assigned to the combination
arm, and $397$ patients were assigned to the chemotherapy only arm. While there 
was a notable delay in treatment effect in this study, the analysis of this study reported in \cite{hellmann2019}
concluded that the nivolumab plus ipilimumab treatment resulted in improved overall survival when compared with chemotherapy.
In our analysis, we utilized survival outcomes that we reconstructed from the published Kaplan-Meier curves for OS in \cite{hellmann2019}. Due to the resolution of these published images, our reconstructed survival outcomes are unlikely to be exactly the same as those recorded in this study, but the reconstructed survival outcomes reproduce the published Kaplan-Meier curves quite closely. Using the reconstructed outcomes, median OS in the nivolumab plus ipilimumab arm was
$17.3$ months while median OS in the chemotherapy arm was $15.0$ months. While median OS suggests an overall
benefit of the combination therapy, the Kaplan-Meier estimates of OS indicate
a delay in treatment effect as the estimated OS survival curve for the chemotherapy arm initially dominates the estimated OS survival curve for the combination therapy arm, and a crossing appears to occur some time between 6 and 9 months before the two Kaplan-Meier estimates clearly separate at later time points.

\begin{figure}
\centering
     \includegraphics[width=5.5in,height=4.0in]{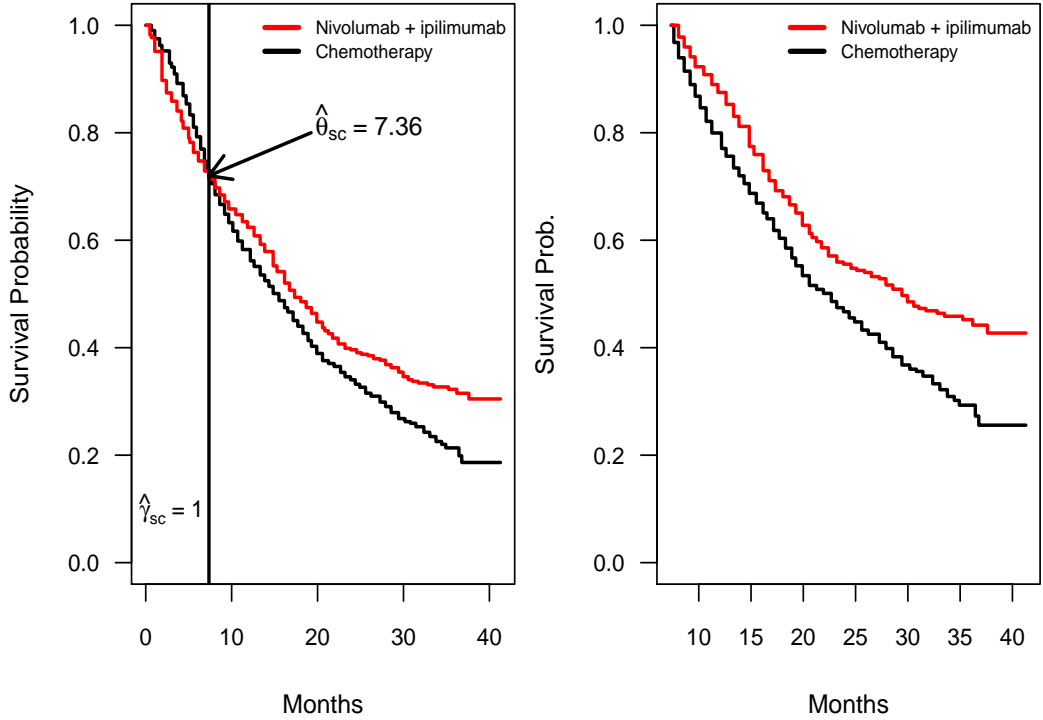}
\caption{Single-crossing constrained estimates of survival in the 
reconstructed data from the nivolumab+iplimumab vs. chemotherapy trial. The left-hand
panel shows the arm-specific survival curve estimates $\hat{S}_{1}^{sc}(t)$ and $\hat{S}_{0}^{sc}(t)$ along with estimates $(\hat{\theta}_{sc}, \hat{\gamma}_{sc})$ of the crossing-time parameters. The right-hand panel shows estimates $\hat{S}_{1,cond}^{sc}(t)$ and $\hat{S}_{0,cond}^{sc}(t)$of the crossing-time conditional survival curves; the conditional survival curves represent survival times conditional on surviving up to the crossing time. }
\label{fig:survival_fitted}
\end{figure}

Figure \ref{fig:survival_fitted} displays the single-crossing constrained estimates of the combination-arm and chemotherapy-arm survival curves for OS. As shown in this figure, the single-crossing constrained survival curve estimate for the chemotherapy arm shows an earlier superiority over the combination-arm survival curve, while the combination-arm survival curve remains superior after the crossing occurs.
The single-crossing constrained estimate $\hat{\theta}_{sc}$ of the crossing time was $\hat{\theta}_{sc} = 7.36$ months, and the corresponding estimate of the initial dominance parameter was $\hat{\gamma}_{sc} = 1$. The right-hand panel of Figure \ref{fig:survival_fitted} shows the single-crossing constrained estimates of the conditional survival curves $S_{a,cond}(t)$ defined in (\ref{eq:conditional_survival_def}). These curves represent estimates of survival probabilities conditional on the fact that one has survived up to the crossing time. The graph of $\hat{S}_{0,cond}^{sc}(t)$ and $\hat{S}_{1,cond}^{sc}(t)$ shows a clear superiority of the active treatment arm among those patients who will survive up to approximately seven and a half months. Indeed, the probability for surviving more than 2 years conditional on surviving up to the crossing time is $0.54$ in the combination arm and $0.47$ in the chemotherapy arm, and the probability for surviving more than 3 years conditional on surviving up to the crossing is $0.44$ in the combination arm and $0.29$ in the chemotherapy arm.

Table \ref{tab:efficacy_summaries} displays single-crossing constrained estimates and their associated $95\%$ confidence intervals for other measures of treatment efficacy. To obtain these confidence intervals, we used a bootstrap with stratified resampling (\cite{davison1997}) where, in each bootstrap replication, a subsample of the survival outcomes $(Y_{i},\delta_{i})$ was drawn with replacement from each of the treatment arms. As shown in this table, our estimate of the proportion surviving up to crossing parameter $S_{a}(\theta)$ was $0.73$ suggesting that approximately $73\%$ of individuals in either treatment arm will survive up to the time point where the active treatment arm will begin to have superior survival probabilities. The estimated difference in RMST truncated at 3 years was $1.48$ months. The estimated difference between the parameters $\textrm{RRML}_{1}(\theta, 36)$ and $\textrm{RRML}_{0}(\theta, 36)$ was $2.43$ months which indicates that, conditional on surviving up to the crossing time, the expected gain in survival time was roughly two and a half months over the time period which begins at the crossing time and ends at 3 years.

\begin{table}[ht]
\centering
\begin{tabular}{crrr}
  \hline
Parameter & Estimate & 2.5\% & 97.5\% \\ 
  \hline
$\theta$ & 7.36 & 4.15 & 23.81 \\
$S_{a}(\theta)$ & 0.73 & 0.37 & 0.86 \\
$\textrm{RMST}_{1}(36) - \textrm{RMST}_{0}(36)$ & 1.48 & -0.50 & 3.33 \\
$\textrm{RRML}_{1}(\theta, 36) - \textrm{RRML}_{0}(\theta, 36)$ & 2.43 & 1.14 & 4.99 \\
$S_{1}(6) - S_{0}(6)$ & -0.03 & -0.09 & 0.02 \\
$S_{1}(12) - S_{0}(12)$ & 0.04 & -0.03 & 0.12 \\
$S_{1}(24) - S_{0}(24)$ & 0.06 & -0.01 & 0.13 \\
$S_{1}(36) - S_{0}(36)$ & 0.11 & 0.05 & 0.18 \\
$S_{1,cond}(12) - S_{0,cond}(12)$ & 0.06 & 0.00 & 0.15 \\
$S_{1,cond}(24) - S_{0,cond}(24)$ & 0.08 & 0.00 & 0.16 \\
$S_{1,cond}(36) - S_{0,cond}(36)$ & 0.15 & 0.08 & 0.24 \\
   \hline
\end{tabular}
\caption{Single-crossing constrained estimates of different
efficacy measures from the reconstructed nivolumab+iplimumab vs. chemotherapy trial.}
\label{tab:efficacy_summaries}
\end{table}

\begin{figure}
\centering
     \includegraphics[width=5.5in,height=4.1in]{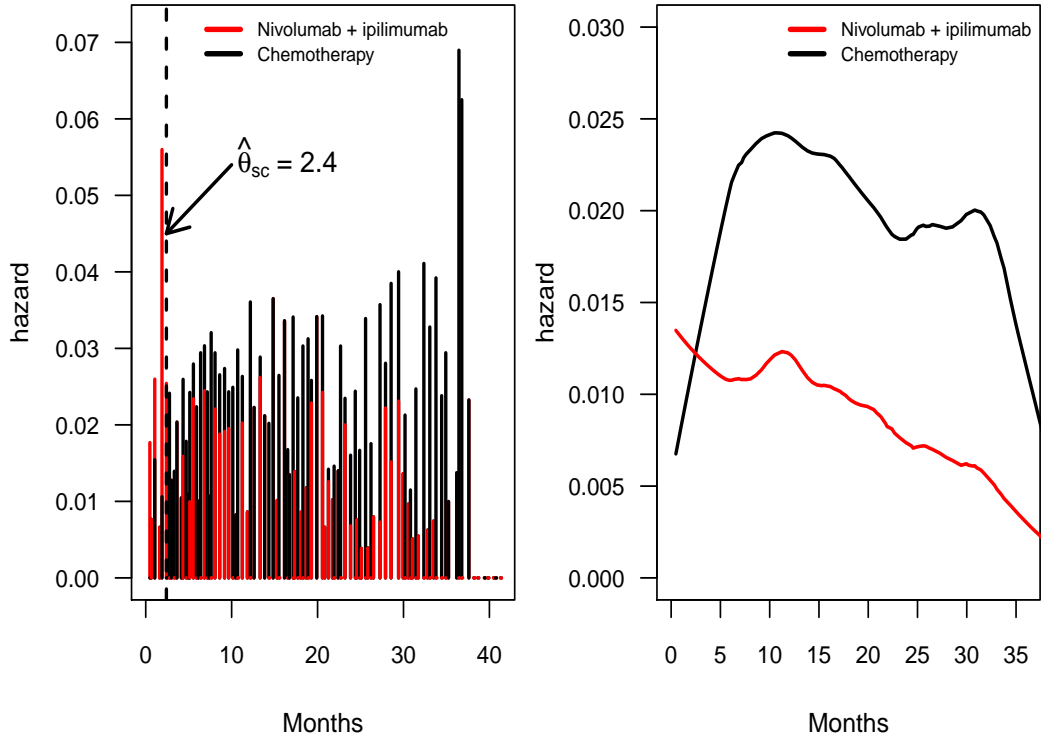}
\caption{Single-crossing constrained estimates of the hazard functions in the 
reconstructed data from the nivolumab+iplimumab vs. chemotherapy trial. The left-hand
panel shows estimates of the arm-specific discrete hazards $h_{ja}(\theta,\gamma)$
along with the estimate of $2.4$ months for the crossing time for the discrete hazards.
The right-hand panel shows the smoothed hazard function for each treatment arm.}
\label{fig:hazard_fitted}
\end{figure}

We also computed estimates of crossing-time parameters and the arm-specific hazard functions under a single-crossing constraint on the hazards rather than the survival curves. Here, we used the approach described in Section \ref{ss:discrete_hazards} where a single-crossing constraint was placed on the discrete hazards with the support of the discrete hazards being placed on the set of observed event times. The left-hand panel of Figure \ref{fig:hazard_fitted} shows the estimated discrete hazards for both treatment arms with the estimated hazard-crossing time of $2.4$ months. This crossing-time estimate suggests that, while those in the combination arm initially have a larger hazard than those in the control arm, the advantage in hazard disappears roughly two and a half months before the hazards actually cross. Using the crossing time of $2.4$ months, estimates of the pre- and post-crossing average hazard ratio parameters described in (\ref{eq:avghaz_parameters}) were $0.77$ and $0.32$ respectively.

While the hazard-based estimate of the crossing time can be useful, the discrete hazards estimates are very non-smooth and hard to interpret. The right-hand panel of Figure \ref{fig:hazard_fitted} shows hazard function estimates obtained by smoothing the discrete hazard estimates in the left-hand panel. To smooth the discrete hazards, we used the LOWESS smoother (\cite{cleveland1979}) with the smoother span set to $2/3$.
We did not impose any additional single-crossing constraints when performing this smoothing, and for the time interval of 0 to 3 years, the single-crossing constraint for the smoothed hazard functions was satisfied without requiring the use of additional constraints on the smoothed functions.

\vspace{-.2in}

\section{Conclusion} \label{sec:discussion}
In this article, we have proposed nonparametric estimators of
two survival curves when such curves are constrained to 
cross at most once. 
The development of these single-crossing constrained estimators was primarily
motivated by clinical trials involving recent cancer immunotherapies 
where it is common to observe delays in treatment effect.
While allowing for more than one crossing could provide additional flexibility, our experience 
with immuno-oncology trials suggests that most successful therapies have at most one
distinct crossing, and cases where one could argue that multiple crossings are present
in the underlying survival curves rarely provide clear evidence of long-term benefit to patients.
Though our approach can improve estimation
performance in cases where the underlying survival curves
conform to a single-crossing constraint, one of the main
advantages of our approach is that it
directly allows for inference on a number of interpretable and useful measures of treatment efficacy. These include the crossing time itself, the proportion of patients who survive past the crossing time, and crossing-time conditional survival probabilities.
When combined with more traditional measures of efficacy, measures such as these can provide important additional context about the benefits or tradeoffs surrounding the active treatment. In addition to estimation with 
single-crossing constraints on the survival functions, we also explored similar nonparametric estimators 
under single-crossing constraints on the hazard functions. Such constraints may be more plausible in
many contexts, and certain efficacy measures such as pre- and post-crossing average hazard ratios
may be more interpretable under single-crossing constraints on the hazard functions.

Though not explored in the present work, the single-crossing constrained estimates
of the crossing parameters could potentially be deployed in the context
of an overall test of the equality of the arm-specific survival curves $S_{0}(t)$ and $S_{1}(t)$.
This could potentially improve power in cases where the active treatment shows a delayed treatment effect and where it is difficult to pre-specify the extent of the delay in treatment effect. 
One possible testing approach is to use a weighted Kaplan-Meier test statistic (\cite{pepe1989}) with a weight function that is only positive at time points after the estimated crossing time.
This would closely resemble the test statistic proposed by \cite{Logan2008} who used a pre-specified rather than estimated time point to determine the support of their weight function. Another attractive alternative would be to consider a weighted log-rank test with a piecewise constant weight function similar to the one proposed in 
\cite{xu2017} and, using the single-crossing constraint on the hazards, specify the jump of the weight function to occur at the estimated crossing time of the hazards. Though establishing the asymptotic null distribution of either the weighted Kaplan-Meier or weighted log-rank test statistic may be challenging, Monte Carlo permutation tests could be used to estimate the desired p-values. 
Even if the single-crossing constrained estimates are not used in constructing a test for comparing the arm-specific survival curves, another use of the single-crossing constrained estimates of the crossing parameters is in the design stage of a study. If relevant historical data are available, one could compute estimates of the crossing time of the hazards or survival functions and such estimates could be used to better inform parameter choices used in sample size and power calculations.

\vspace{-.2in}
\section*{Supplemental Information}
An \verb"R" package \verb"DelayedSurvFit" implementing the methods described in this article and
containing the reconstructed dataset analyzed in Section 5 is publicly
available at \url{https://github.com/nchenderson/DelayedSurvFit}.
The \verb"R" code used to conduct the simulation study described in Section 4
and the \verb"R" code used for the data analysis shown in Section 5 are available at
\url{https://github.com/nchenderson/singlecrossingreproduce}.
\vspace{-.1in}

\bibliographystyle{biorefs}
\bibliography{delayed_bib}

\begin{thebibliography}{99}

\bibitem[Boyd and Vandenberghe(2004)Boyd and Vandenberghe]{boyd2004convex}
\textsc{Boyd, Stephen and Vandenberghe, Lieven}. (2004).
\newblock {\em Convex optimization\/}. Cambridge university press.

\bibitem[Chen(2013)Chen]{chen2013}
\textsc{Chen, Tai-Tsang}. (2013).
\newblock Statistical issues and challenges in immuno-oncology.
\newblock {\em Journal for immunotherapy of cancer\/}~\textbf{1}(1), 18.

\bibitem[Chen(2015)Chen]{chen2015}
\textsc{Chen, Tai-Tsang}. (2015).
\newblock Milestone survival: a potential intermediate endpoint for immune
  checkpoint inhibitors.
\newblock {\em Journal of the National Cancer Institute\/}~\textbf{107}(9),
  djv156.

\bibitem[Cleveland(1979)Cleveland]{cleveland1979}
\textsc{Cleveland, William~S}. (1979).
\newblock Robust locally weighted regression and smoothing scatterplots.
\newblock {\em Journal of the American statistical
  association\/}~\textbf{74}(368), 829--836.

\bibitem[Cortese \emph{and others}(2017)Cortese, Holmboe and
  Scheike]{cortese2017}
\textsc{Cortese, Giuliana, Holmboe, Stine~A and Scheike, Thomas~H}. (2017).
\newblock Regression models for the restricted residual mean life for
  right-censored and left-truncated data.
\newblock {\em Statistics in medicine\/}~\textbf{36}(11), 1803--1822.

\bibitem[Davison and Hinkley(1997)Davison and Hinkley]{davison1997}
\textsc{Davison, Anthony~Christopher and Hinkley, David~Victor}. (1997).
\newblock {\em Bootstrap methods and their application\/}, Number~1. Cambridge
  university press.

\bibitem[Demarqui and Mayrink(2019)Demarqui and Mayrink]{demarqui2019}
\textsc{Demarqui, Fabio~N and Mayrink, Vinicius~D}. (2019).
\newblock A fully likelihood-based approach to model survival data with
  crossing survival curves.
\newblock {\em arXiv:1910.02406\/}.

\bibitem[Harrington and Fleming(1982)Harrington and Fleming]{harrington1982}
\textsc{Harrington, David~P and Fleming, Thomas~R}. (1982).
\newblock A class of rank test procedures for censored survival data.
\newblock {\em Biometrika\/}~\textbf{69}(3), 553--566.

\bibitem[Hellmann \emph{and others}(2019)Hellmann, Paz-Ares, Bernabe~Caro,
  Zurawski, Kim, Carcereny~Costa, Park, Alexandru, Lupinacci, de~la
  Mora~Jimenez  et~al.]{hellmann2019}
\textsc{Hellmann, Matthew~D, Paz-Ares, Luis, Bernabe~Caro, Reyes, Zurawski,
  Bogdan, Kim, Sang-We, Carcereny~Costa, Enric, Park, Keunchil, Alexandru,
  Aurelia, Lupinacci, Lorena, de~la Mora~Jimenez, Emmanuel  \emph{and others}}.
  (2019).
\newblock Nivolumab plus ipilimumab in advanced non--small-cell lung cancer.
\newblock {\em New England Journal of Medicine\/}~\textbf{381}(21), 2020--2031.

\bibitem[Johansen(1978)Johansen]{johansen1978}
\textsc{Johansen, S{\o}ren}. (1978).
\newblock The product limit estimator as maximum likelihood estimator.
\newblock {\em Scandinavian Journal of Statistics\/}~\textbf{5}(4), 195--199.

\bibitem[Kalbfleisch and Prentice(1981)Kalbfleisch and
  Prentice]{kalbfleisch1981}
\textsc{Kalbfleisch, John~D and Prentice, Ross~L}. (1981).
\newblock Estimation of the average hazard ratio.
\newblock {\em Biometrika\/}~\textbf{68}(1), 105--112.

\bibitem[Lawless(2011)Lawless]{lawless2011}
\textsc{Lawless, Jerald~F}. (2011).
\newblock {\em Statistical models and methods for lifetime data\/}, Volume 362.
  John Wiley \& Sons.

\bibitem[Lin \emph{and others}(2020)Lin, Lin, Roychoudhury, Anderson, Hu,
  Huang, Leon, Liao, Liu, Luo  et~al.]{lin2020}
\textsc{Lin, Ray~S, Lin, Ji, Roychoudhury, Satrajit, Anderson, Keaven~M, Hu,
  Tianle, Huang, Bo, Leon, Larry~F, Liao, Jason~JZ, Liu, Rong, Luo, Xiaodong
  \emph{and others}}. (2020).
\newblock Alternative analysis methods for time to event endpoints under
  nonproportional hazards: A comparative analysis.
\newblock {\em Statistics in Biopharmaceutical Research\/}~\textbf{12}(2),
  187--198.

\bibitem[Logan \emph{and others}(2008)Logan, Klein and Zhang]{Logan2008}
\textsc{Logan, Brent~R, Klein, John~P and Zhang, Mei-Jie}. (2008).
\newblock Comparing treatments in the presence of crossing survival curves: an
  application to bone marrow transplantation.
\newblock {\em Biometrics\/}~\textbf{64}(3), 733--740.

\bibitem[Nocedal and Wright(2006)Nocedal and Wright]{nocedal2006}
\textsc{Nocedal, Jorge and Wright, Stephen}. (2006).
\newblock {\em Numerical optimization\/}. Springer Science \& Business Media.

\bibitem[Owen(2001)Owen]{owen2001}
\textsc{Owen, Art~B}. (2001).
\newblock {\em Empirical likelihood\/}. Chapman and Hall/CRC.

\bibitem[Pak \emph{and others}(2017)Pak, Uno, Kim, Tian, Kane, Takeuchi, Fu,
  Claggett and Wei]{pak2017}
\textsc{Pak, Kyongsun, Uno, Hajime, Kim, Dae~Hyun, Tian, Lu, Kane, Robert~C,
  Takeuchi, Masahiro, Fu, Haoda, Claggett, Brian and Wei, Lee-Jen}. (2017).
\newblock Interpretability of cancer clinical trial results using restricted
  mean survival time as an alternative to the hazard ratio.
\newblock {\em JAMA oncology\/}~\textbf{3}(12), 1692--1696.

\bibitem[Park \emph{and others}(2012)Park, Kalbfleisch and Taylor]{park2012}
\textsc{Park, Yongseok, Kalbfleisch, John~D and Taylor, Jeremy~MG}. (2012).
\newblock Constrained nonparametric maximum likelihood estimation of
  stochastically ordered survivor functions.
\newblock {\em Canadian Journal of Statistics\/}~\textbf{40}(1), 22--39.

\bibitem[Pepe and Fleming(1989)Pepe and Fleming]{pepe1989}
\textsc{Pepe, Margaret~Sullivan and Fleming, Thomas~R}. (1989).
\newblock Weighted {K}aplan-{M}eier statistics: a class of distance tests for
  censored survival data.
\newblock {\em Biometrics\/}~\textbf{45}, 497--507.

\bibitem[Rahman \emph{and others}(2019)Rahman, Fell, Ventz, Arf{\'e},
  Vanderbeek, Trippa and Alexander]{rahman2019}
\textsc{Rahman, Rifaquat, Fell, Geoffrey, Ventz, Steffen, Arf{\'e}, Andrea,
  Vanderbeek, Alyssa~M, Trippa, Lorenzo and Alexander, Brian~M}. (2019).
\newblock Deviation from the proportional hazards assumption in randomized
  phase 3 clinical trials in oncology: prevalence, associated factors, and
  implications.
\newblock {\em Clinical Cancer Research\/}~\textbf{25}(21), 6339--6345.

\bibitem[Royston and Parmar(2013)Royston and Parmar]{royston2013}
\textsc{Royston, Patrick and Parmar, Mahesh~KB}. (2013).
\newblock Restricted mean survival time: an alternative to the hazard ratio for
  the design and analysis of randomized trials with a time-to-event outcome.
\newblock {\em BMC medical research methodology\/}~\textbf{13}(1), 152.

\bibitem[Schemper \emph{and others}(2009)Schemper, Wakounig and
  Heinze]{schemper2009}
\textsc{Schemper, Michael, Wakounig, Samo and Heinze, Georg}. (2009).
\newblock The estimation of average hazard ratios by weighted {C}ox regression.
\newblock {\em Statistics in medicine\/}~\textbf{28}(19), 2473--2489.

\bibitem[Xu \emph{and others}(2017)Xu, Zhen, Park and Zhu]{xu2017}
\textsc{Xu, Zhenzhen, Zhen, Boguang, Park, Yongsoek and Zhu, Bin}. (2017).
\newblock Designing therapeutic cancer vaccine trials with delayed treatment
  effect.
\newblock {\em Statistics in medicine\/}~\textbf{36}(4), 592--605.

\bibitem[Yang and Prentice(2005)Yang and Prentice]{yang2005}
\textsc{Yang, Song and Prentice, Ross}. (2005).
\newblock Semiparametric analysis of short-term and long-term hazard ratios
  with two-sample survival data.
\newblock {\em Biometrika\/}~\textbf{92}(1), 1--17.

\bibitem[Zhao \emph{and others}(2016)Zhao, Claggett, Tian, Uno, Pfeffer,
  Solomon, Trippa and Wei]{zhao2016}
\textsc{Zhao, Lihui, Claggett, Brian, Tian, Lu, Uno, Hajime, Pfeffer, Marc~A,
  Solomon, Scott~D, Trippa, Lorenzo and Wei, LJ}. (2016).
\newblock On the restricted mean survival time curve in survival analysis.
\newblock {\em Biometrics\/}~\textbf{72}(1), 215--221.

\bibitem[Zhou(2015)Zhou]{zhou2015}
\textsc{Zhou, Mai}. (2015).
\newblock {\em Empirical likelihood method in survival analysis\/}. Chapman and
  Hall/CRC.

\end{thebibliography}

\label{lastpage}

\end{document}